\documentclass[12pt]{article}
\pdfoutput=1
\usepackage[utf8]{inputenc}
\usepackage{amsmath}
\usepackage{amsfonts}
\usepackage{amssymb}
\usepackage{graphicx}
\usepackage[left=2cm,right=2cm,top=4cm,bottom=3cm]{geometry}
\usepackage[space]{grffile}
\usepackage{flafter}
\usepackage{multirow}
\usepackage{tikz}
\usetikzlibrary{shapes,arrows}
\usepackage{epsfig,amsfonts,amsthm}
\usepackage[normalem]{ulem}
\usepackage{amsmath,amssymb}
\usepackage{array}
\usepackage{amsmath}
\usepackage{amsfonts}
\usepackage{amssymb}
\usepackage{subfig}
\usepackage{wrapfig}
\usepackage{graphicx}
\usepackage{cite}
\usepackage{hyperref}
\newcommand{\be}{\begin{equation}}
\newcommand{\ee}{\end{equation}}
\newcommand{\bea}{\begin{eqnarray}}
\newcommand{\eea}{\end{eqnarray}}

\usepackage{color}

\def\lsim{\mathrel{\rlap{\lower4pt\hbox{\hskip1pt$\sim$}}
    \raise1pt\hbox{$<$}}}         
\def\gsim{\mathrel{\rlap{\lower4pt\hbox{\hskip1pt$\sim$}}
    \raise1pt\hbox{$>$}}}         

\def\beq{\begin{equation}}
\def\eeq{\end{equation}}
\def\bea{\begin{eqnarray}}
\def\eea{\end{eqnarray}}

\def\<{\left\langle}
\def\>{\right\rangle}

\usepackage[normalem]{ulem}

\def\lsim{\mathrel{\rlap{\lower4pt\hbox{\hskip1pt$\sim$}}
    \raise1pt\hbox{$<$}}}         
\def\gsim{\mathrel{\rlap{\lower4pt\hbox{\hskip1pt$\sim$}}
    \raise1pt\hbox{$>$}}}         

\def\beq{\begin{equation}}
\def\eeq{\end{equation}}
\def\bea{\begin{eqnarray}}
\def\eea{\end{eqnarray}}

\def\<{\left\langle}
\def\>{\right\rangle}

\newcommand{\bt}{\begin{tabular}}
\newcommand{\et}{\end{tabular}}

\usepackage{hyperref}

\hypersetup{
   colorlinks=true,       
   linkcolor=blue,        
   citecolor=red,         
   filecolor=magenta,      
   urlcolor=cyan,           
   linktocpage = true,
   }

\usepackage{tikz}
\usetikzlibrary{decorations.pathmorphing,decorations.markings}

\usepackage{graphicx}

\tikzset{
photon/.style={decorate, decoration={snake,amplitude=2pt, segment length=5pt}, draw=black},
particle/.style={draw=black, postaction={decorate}, decoration={markings,mark=at position .5 with {\arrow[draw=black]{>}}}},
antiparticle/.style={draw=black, postaction={decorate}, decoration={markings,mark=at position .5 with {\arrowreversed[draw=black]{>}}}},
gluon/.style={decorate, draw=black, decoration={coil,amplitude=4pt, segment length=5pt}},
goldstone/.style={draw=green,postaction={decorate},decoration={markings,mark=at position .5 with {\arrow[draw=blue]{>}}}}
}

\begin{document}

\title{\hfill ~\\[-50mm]
                  \textbf{\large Fat $b$-Jet Analyses Using Old and New Clustering Algorithms \\
in New Higgs Boson Searches at the LHC
                }        }
\date{}

\author{\\[-5mm]
A. Chakraborty\footnote{E-mail: {\small\tt amit.c@srmap.edu.in}} $^{~1}$,\
S. Dasmahapatra\footnote{E-mail: {\small\tt sd@ecs.soton.ac.uk}} $^{~2}$,\
H.A. Day-Hall\footnote{E-mail: {\small\tt hadh1g17@soton.ac.uk}} $^{~4}$,\
B. Ford\footnote{E-mail: {\small\tt b.ford@soton.ac.uk}} $^{~3}$,\ \\
S. Jain\footnote{E-mail: {\small\tt s.jain@soton.ac.uk}} $^{~3}$,\
S. Moretti\footnote{E-mail: {\small\tt stefano@soton.ac.uk}}
\footnote{E-mail: {\small\tt stefano.moretti@physics.uu.se}} $^{~~3,5}$
\\ \\
\emph{\small $^1$Department of Physics, School of Engineering and Sciences, SRM University AP,}\\
\emph{\small  Amaravati, Mangalagiri 522240, India}\\
\emph{\small $^2$School of Electronics and Computer Science, University of Southampton,}\\
\emph{\small Southampton, SO17 1BJ, United Kingdom}\\
\emph{\small $^3$School of Physics and Astronomy, University of Southampton,}\\
\emph{\small Southampton, SO17 1BJ, United Kingdom}\\
\emph{\small $^4$Faculty of Nuclear Sciences and Physical Engineering, Czech Technical University in Prague,}\\
\emph{\small Brehova 78/7, 11519 Stare Mesto, Czechia}\\
\emph{\small $^5$Department of Physics and Astronomy, Uppsala University, Uppsala, Sweden}\\
[3mm]
  }

\maketitle

\vspace*{-10mm}

\begin{abstract}
\noindent
{We compare different jet-clustering algorithms in establishing fully hadronic final states stemming from the chain decay of a heavy Higgs state into a pair of the 125 GeV Higgs boson that decays into bottom-antibottom quark pairs. Such $4b$ events typically give rise to boosted topologies, wherein $b\bar b$ pairs emerging from each 125 GeV Higgs boson  tend to merge into a single, fat $b$-jet. Assuming Large Hadron Collider (LHC) settings, we illustrate how both the efficiency of selecting the multi-jet final state and the ability to reconstruct from it the masses of all Higgs bosons depend on the choice of jet-clustering algorithm and its parameter settings.  We indicate the optimal choice of clustering method for the purpose of establishing such a ubiquitous Beyond the SM (BSM) signal, illustrated via a Type-II 2-Higgs Doublet Model (2HDM).  }
 \end{abstract}
\thispagestyle{empty}
\vfill
\newpage
\tableofcontents

\vspace{0.5em}
\section{Introduction}
The Higgs boson discovered in 2012 at the LHC has been extensively studied, so that we now know that it is very SM-like \cite{Aad:2012tfa}. That is, it is clear that its quantum numbers (charge, spin, CP) are consistent with those predicted in the SM and so are its couplings, at least those measured thus far, to $W^\pm$ and $Z$ bosons as well  as to $t,b,c,\tau$ and $\mu$ fermions.  Amongst of all these, the $Hb\bar b$ coupling plays a particular role, for a twofold reason. On the one hand, it is the dominant one for the SM-like Higgs state,
as the $b\bar b$ decay rate is the largest \cite{Moretti:1994ds,Djouadi:1995gv},  while also being the one most  
polluted by large backgrounds (chiefly including the overwhelming $t\bar t$ production and decay). On the other hand, the $b\bar b$ decay channel  presents significant challenges experimentally, primarily connected to the necessity of flavour tagging it amongst myriads of light-quark and gluon jets stemming from the majority of Quantum Chromo-Dynamics  (QCD) interactions, apart from $b$-jets from background $t\bar t$ decays. 

It is therefore important to assess the current status of phenomenological approaches to the extraction of these multi-$b$-jet signals. While there exists copious literature on this topic within the SM, wherein, in the foreseeable future (i.e., Run 3 of the LHC), the SM-like Higgs state can only be produced singly\footnote{In fact, di-Higgs  production within the SM will only become accessible at the High-Luminosity LHC (HL-LHC) \cite{Gianotti:2002xx}.}, comparatively less developed are studies of pair production in Beyond the SM (BSM) scenarios, despite significant cross sections for a resonant decay of a heavier Higgs boson leading to the Higgs  pair decaying to a $4b$ final state. This is why, in Ref.~\cite{Chakraborty:2020vwj}, we studied the process $pp\to H\to hh\to 4b$, for $m_H=125$ GeV and $m_h$ between 40 and 60 GeV, which would be a striking signal of, for example, a 2-Higgs Doublet Model (2HDM) \cite{Gunion:1989we, Gunion:1992hs, Branco:2011iw} in the so-called `inverted hierarchy' scenario, i.e., when the discovered Higgs state is not the lightest one. 

In that paper, we assessed the ability of different jet-clustering algorithms, with different resolution parameters and reconstruction procedures, to resolve such fully hadronic final states. Therein, we showed that both the efficiency of selecting the hadronic states and the ability to reconstruct Higgs masses from these depend strongly on the choice of  the jet-clustering algorithm and its settings. Specifically, we emphasised that variable-$R$ algorithms \cite{Krohn:2009zg} were more effective in gaining signal sensitivity as well as in reconstructing the light and heavy Higgs mass peaks, than those based on a fixed cone radius $R$ \cite{Cacciari:2008gp, Salam:2009jx}.

Those results were obtained for slim $b$-jets, for which no merging occurred (so we looked at typical four $b$-jet configurations). In the present paper, we want to instead study the case of fat $b$-jets, i.e., when two $b$-partons emerging from a $h$ decay are not resolved as individual jets, but are merged into a fat $b$-jet containing both. This is most likely to occur when the $H$ state is significantly heavier than the $h$, $m_H\gg m_h=125$  in the usual 2HDM in the `standard hierarchy' scenario. Again, we will assess which of the two types of jet-clustering algorithms, fixed or variable cone size, is better able to extract the signal from the backgrounds and yield the sharpest rendition of the  Breit-Wigner  mass peaks. For this purpose, we will implement a simplified (MC truth informed) double $b$-tagger. It is worthwhile to mention that one can use other boosted jet tagging methods based on the jet substructure technique to further enhance the signal significances, e.g., N-subjettiness variables and their ratios \cite{Thaler:2010tr}, Energy Correlation Functions (ECF) and their ratios \cite{Chakraborty:2020yfc,Bhattacherjee:2022gjq}, or a combination of substructure based observables and cutting edge machine learning techniques \cite{Larkoski:2017jix}. Many experimental studies have been done on the fat jets analysis \cite{ATLAS:2019qdc,ATLAS:2018rnh,CMS:2017aza,CMS:2016jvt}.

The plan of the paper is as follows. In the next section, we  describe how jets are defined at the LHC. We then move on to describe the Monte Carlo (MC) analysis that we will perform (i.e., simulation tools, cutflow, $b$-tagger, etc.). After which we will present our results. Finally, in the last section, we will draw our conclusions.

\section{Jets at the LHC}
In a modern particle collider, such as the LHC, the most crucial difficulty in extracting new physics is making sense out of the mess of particles collected in the detectors in each event. A so-called jet definition provides a mapping between hard interactions in our
Quantum Field Theory (QFT), which is what we are ultimately looking to test, and the jumble of particles we actually observe in the detectors.

One of the well-known peculiarities of QCD is colour confinement, i.e., the fact that quarks and gluons cannot exist as free particles, instead only appear inside hadronic bound states. In the high energy environment of the LHC, they undergo showering and hadronisation and are  detected as sprays of (colourless) hadrons.

A simple, intuitive picture of this process is to consider the emission rate for a quark(antiquark) to radiate a gluon, given by \cite{Gribov:1972rt,Gribov:1972ri,Altarelli:1977zs,Dokshitzer:1977sg}
\begin{equation}
\mathcal{P}_{gq}(z) = C_F\biggr[\frac{1+(1-z)^2}{z} \biggr],
\end{equation}
and, similarly, for a gluon radiating another gluon,
\begin{flalign}
\nonumber
\mathcal{P}_{gg}(z) =& C_A \biggr[\frac{z}{1-z}+\frac{1-z}{z}+z(1-z)\biggr]+
\\
&+\delta(1-z)\frac{(11 C_A-4 n_f T_R)}{6},
\end{flalign}
Sometimes, a gluon could also split into a quark-antiquark pair, according to
 \begin{equation}
\mathcal{P}_{qg}(z)= T_R \biggr[z^2+(1-z)^2\biggr],
\end{equation}
where $z$ and $(1-z)$ are the energy fractions, $n_f$ is the number of fermions coupling to the gluons, $C_F$, $C_A$ and $T_R$ are the usual QCD `colour factors'.

These splittings repeat themselves in all possible combinations,  thereby generating the aforementioned shower, wherein partons are rather soft and/or collinear (note the $E$ and $\theta$ `divergences'), so that the final partons are rather collimated in the direction of the primary ones. Once the energy of the initial collision is spread amongst all these subsequent partons so that the average value of it is close to $\Lambda_{\rm QCD}$, the hadronisation process takes place by generating hadrons (pions, kaons, etc.) which directions are also aligned with those of the primary partons (assuming that $Q\gg \Lambda_{\rm QCD}$). (Recall that, owing to the running of the QCD coupling constant,  
the partonic couplings will reach the non-perturbative regime before partons reach the detectors.) The end result is the creation of the aforementioned sprays of hadrons, called jets. However, no matter how intuitive this qualitative picture is, one needs a quantitative algorithm to define such jets.
\subsection{Jet Clustering Algorithms}

We here review two classes of jet clustering algorithms currently in use at the LHC.

\subsubsection{Fixed Cone Jets}
To provide a mapping between hard interactions and the hadronic sprays observed in particle detectors,  algorithmic procedures are used to characterise the aforementioned jets. Over the years, there has been extensive development of jet clustering algorithm, beginning in 1977 with Sterman and Weinberg \cite{Sterman:1977wj}, who  indeed defined jets as cones, initially deployed in the context of $e^+ e^- \rightarrow$ hadron scatterings. The type of algorithms currently employed at the LHC, and of particular interest for this study, are known as sequential recombination algorithms \cite{Moretti:1998qx}, or ‘jet clustering algorithms’.

Jet clustering algorithms reduce the complexity of final states by attempting to rewind the showering/hadronisation process. They  consider each particle in an event and all are iteratively combined together based on some inter-particle distance measure to form jets. Remarkably, when a jet clustering algorithm is well designed, it can be applied at both the parton and hadron levels, so as to enable one to make direct comparisons between theory and experiment.

All (sequential) jet clustering algorithms currently used at the LHC employ a similar method descending from a generalised  $k_T$ algorithm. This uses an inter-particle distance measure between two particles ($i$ and $j$), given as
\begin{equation}
 d_{ij}={\rm min}(p_{Ti}^{2n},p_{Tj}^{2n})\frac{\Delta R^2_{ij}}{R^2},
 \end{equation}
 where $\Delta R^2_{ij}=(y_i-y_j)^2+(\phi_i-\phi_j)^2$ is an angular distance between two particles $i$ and $j$, with $y$ and $\phi$ being the rapidity and azimuth of the associated final state hadron, $n$ is an exponent corresponding to a particular jet clustering algorithm and $R$ is the jet radius (or cone radius) parameter. 
The second distance variable is the ‘beam distance’: 
 \begin{equation}
     d_{Bi}=p_{Ti}^{2n},
 \end{equation}
 which is the separation between object $i$ and the beam $B$. The algorithm works by finding the minimum $d_{\rm min}$ of all the $d_{ij}$'s and $d_{Bi}$'s and then the following happens. 
 \begin{itemize}
     \item If $d_{\rm min}$ is a $d_{ij}$, combine $i$ and $j$ then repeat the process.
     \item If $d_{\rm min}$ is a $d_{Bi}$, then $i$ is declared a jet and removed from the list. This procedure is then repeated until no particles are left.
\end{itemize}
If we now take into account some pair of pseudojets $i$ and $j$, with $i$ having lower $p_T$ than $j$ and being selected in $d_{ij}$, we can write (for $n\ge0$)
\begin{equation}
{d_{ij} = \frac{\Delta R^2_{ij}}{R^2} p^{2n}_{Ti} = \frac{\Delta R^2_{ij}}{R^2} d_{Bi}.}
\end{equation}
For $n\le0$, it will be other way around with $p_{Tj}$ selected from $d_{ij}$ and the above equation will change to:
\begin{equation}
{d_{ij} = \frac{\Delta R^2_{ij}}{R^2} p^{2n}_{Tj} = \frac{\Delta R^2_{ij}}{R^2} d_{Bj}.}
\end{equation}

We require the ratio $\frac{\Delta R^2_{ij}}{R^2}<1$ to evade declaring $i$ a jet instead of combining $i$ with $j$, therefore parameter $R$ acts as a cut-off for the particle pairing and is proportional to the final size of the jets.
 The algorithms currently most  used at the LHC are the anti-$k_T$ \cite{Cacciari:2008gp, Salam:2009jx} and the Cambridge/Aachen (C/A) ones \cite{Wobisch:1998wt, Dokshitzer:1997in}. The $n$ value for the anti-$k_T$ and C/A algorithms are $-1$ and 0, respectively. 
 
\subsubsection{Variable-$\boldmath{R}$ Jets}
The fixed input parameter, $R$, mentioned above acts as a cut-off for particle pairing  and applying a size limit on the jets based on the separation between the particles. We also know that the angular spread of the jet constituents depends on the initial partons $p_T$. For objects with high $p_T$, the decay products are more tightly packed into a collimated cone whereas for the objects with lower $p_T$, the constituents will be spread over some wider angle. Therefore, it is very important to carefully select the right $R$ value for clustering  depending on the relevant $p_T$ distribution to capture the underlying physics.
  
  The more recent variation of the standard jet clustering algorithms is the so-called Variable-$R$ jet clustering algorithm \cite{Krohn:2009zg}, which alters the scheme mentioned in Section 2.1.1 to adapt with jets of varying cone size in an event. A modification is made to the distance measure $d_{ij}$, such that:
\begin{equation}
d_{ij}={\rm min}(p_{Ti}^{2n},p_{Tj}^{2n})\Delta R^2_{ij}
 \end{equation}
 and
\begin{equation}
d_{Bi}=p_{Ti}^{2n} R^2.
 \end{equation}
 Next, the  fixed input parameter $R$ is replaced by a $p_T$ dependent $R_{\rm eff}(p_{Ti})=\frac{\rho}{p_T}$, where $\rho$ is a dimensionful input parameter, such that:
 \begin{equation}
d_{Bi}=p_{Ti}^{2n} R_{\rm eff}(p_{Ti})^2.
 \end{equation}
 For objects with larger $p_T$,  $d_{Bi}$ will be suppressed and these objects are more likely to be classified as jets, For objects with lower $p_T$, these are combined with the nearest neighbour increasing the spread of constituents as $d_{Bi}$ is more enhanced.

In the variable-$R$ approach, the process is modified in such a way that one can avoid events with very wide jets at low $p_T$. The dimensionful parameter $\rho$ can be scanned over a range to optimise the maximum desired sensitivity. This can also be done for other parameters such as $R_{\rm min/max}$ (cut-offs for the minimum and maximum allowed $R_{\text{\rm eff}}$),
respectively,  i.e., if a jet has $R_{\text{\rm eff}} < R_{\rm min}$, it is overwritten and set to $R_{\text{\rm eff}} = R_{\rm min}$ and equivalently for $R_{\rm max}$.
%

At last, we hypothesise that using a variable-$R$ clustering procedure can show improvement in reconstructing signal when compared to traditional fixed-$R$ routines. The variable-$R$ technique helps in reducing the complexity of finding a suitable single fixed cone size to envelop all the radiation without including too much outside noise or 'junk' inside a jet. This will be quite useful for our study as we look at constructing fatjets.

\section{Methodology}
In this section, we describe the tools and selection strategy to pursue our analysis.
\subsection{Simulation Details}
We consider a suitable Benchmark Point (BP) in the context of the 
2HDM Type-II (2HDM-II henceforth)  where we assume the lightest CP-even Higgs state to be the SM-like Higgs Boson with $m_h$ = 125 GeV and set the heavier CP-even Higgs boson mass as $m_H$ = 700 GeV. The BP has been tested against theoretical and experimental constraints by using {\tt 2HDMC} \cite{Eriksson:2010zzb} interfaced with {\tt HiggsBounds} \cite{ Bechtle:2013wla} and {\tt HiggsSignals} \cite{ Bechtle:2013xfa} and against flavour constraints using {\tt SuperISO} \cite{ Mahmoudi:2009zz}. Specifically, concerning the latter, the 
 following flavour constraints on meson decay Branching Ratios (BRs) and mixings, all to the $2\sigma$ level, are used in our analysis: ${\rm BR}(b \rightarrow s \gamma)$, ${\rm BR}(B_s \rightarrow \mu \mu)$, ${\rm BR}(D_s \rightarrow \tau \nu)$, ${\rm BR}(D_s \rightarrow \mu \nu)$, ${\rm BR}(B_u \rightarrow \tau \nu)$, $\frac{{\rm BR}(K \rightarrow \mu\nu)}{{\rm BR}(\pi \rightarrow \mu \nu)}$, ${\rm BR}(B \rightarrow D_0 \tau \nu)$ and $\Delta_0(B \rightarrow K^* \gamma)$. 

Our study assumes proton-proton collisions at a center-of-mass energy of 13 TeV and integrated luminosities of 140 and 300 ${\rm fb^{-1}}$, corresponding to full Run 2 and Run 3 datasets.  The production cross sections at Leading Order (LO)\footnote{This is also the perturbative level at which MC events are generated.}  and decay rates for the sub-processes $gg$, $q\bar q$ $\to H\to hh\to b\bar b b\bar b$
are presented in Tab.~\ref{tab:params}, alongside the 2HDM-II input parameters. In the calculation of the overall cross-section, the  renormalisation and factorisation scales were both set to be $H_{T}/2$, where $H_T$ is the sum of the transverse energy of each parton. The Parton Distribution Function (PDF) set used was ${\tt NNPDF23_{\bf -}lo_{\bf -}as_{\bf -}0130_{\bf -}qed}$ \cite{Ball:2014uwa}. 
Finally, in order to carry out a realistic MC simulation, the toolbox described in Fig.~\ref{fig:toolbox1} was used to generate and analyse events.

\begin{table}[!h]
\begin{center}

\scalebox{0.6}{
\begin{tabular}{ |c|c|c|c|c|c|c|c|c|c| }
 \hline
 Label & $m_h$ (GeV) & $m_H$ (GeV) & $\tan\beta$ & $\sin (\beta -\alpha)$ & $m_{12}^2$ & BR($H\rightarrow hh$) & BR($h\rightarrow b\overline{b}$)&$\sigma$(pb)  \\
 \hline
BP1 & 125 & 700.668 & 2.355& -0.999 & 1.46$\times 10^5$ &  6.218$\times 10^{-1}$ &6.164$\times 10^{-1}$&1.870$\times 10^{-2}$\\

\hline
\end{tabular}
}
\caption{\label{tab:params} The 2HDM-II  parameters and LO cross-section of the process studied for our BP.}
\end{center}
\end{table}

\vspace*{1em}
\tikzstyle{node} = [rectangle, rounded corners, minimum width=3cm, minimum height=1cm,text centered, draw=black]
\tikzstyle{arrow} = [thick,->,>=stealth]

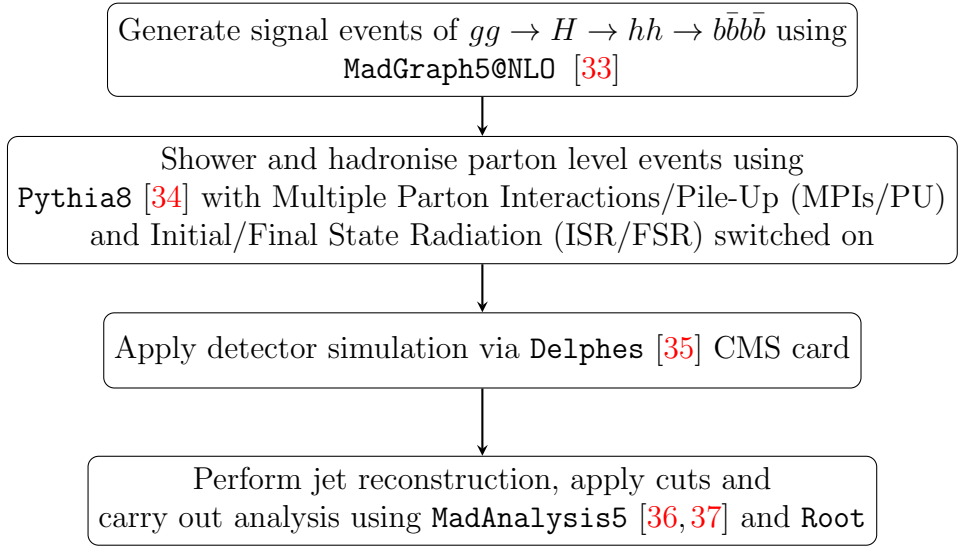
\begin{figure}[!htb]
\centering
\hspace*{-1.5truecm}
\begin{tikzpicture}[node distance=2cm]
\node (step1) [node, align=center] {Generate signal events of $gg \rightarrow H \rightarrow h h \rightarrow b  \bar{b} b\bar{b}$ using \\ {\tt MadGraph5@NLO }\cite{Alwall:2014hca}};
\node (step2) [node, align=center, below of=step1] {Shower and hadronise parton level events using\\ {\tt Pythia8} \cite{Sjostrand:2007gs} with Multiple Parton Interactions/Pile-Up (MPIs/PU)\\ and Initial/Final State Radiation (ISR/FSR) switched on};
\node (step3) [node, align=center, below of=step2] {Apply detector simulation via {\tt Delphes} \cite{deFavereau:2013fsa} CMS card};
\node (step4) [node, align=center, below of=step3] {Perform jet reconstruction, apply cuts and \\ carry out analysis using {\tt MadAnalysis5} \cite{Conte:2012fm,Conte:2018vmg} and {\tt Root}};

\draw [arrow] (step1) -- (step2);
\draw [arrow] (step2) -- (step3);
\draw [arrow] (step3) -- (step4);
\end{tikzpicture}
\caption{Description of the procedure used to generate and analyse MC events. }
\label{fig:toolbox1}
\end{figure}
The same toolkit (see Fig.~\ref{fig:toolbox1}) is used to generate samples of the leading SM backgrounds. The background processes we consider are the following: the QCD $4b$ background, $gg,q \bar q \rightarrow t \bar t$ and  $gg,q \bar q \rightarrow Zb\bar b$ \cite{Chakraborty:2020vwj}. Due to the kinematic differences between the signal process and leading backgrounds, we apply generation level cuts within 
{\tt{MadGraph5}}  to improve the selection efficiency at the jet level, as
follows:
\begin{equation}
\nonumber
 gg,q \bar q \to t \bar t: \ p^{\rm gen}_T (t) > 250\ {\rm GeV},
\end{equation}
\begin{equation}
\nonumber
 gg,q \bar q \to b \bar b b \bar b: \ p^{\rm gen}_T (b) > 100\ {\rm GeV},
\end{equation}
\begin{equation}
\nonumber
 gg,q \bar q \to Z b \bar b: \ p^{\rm gen}_T (Z) > 250\ {\rm GeV}, \ p^{\rm gen}_T (b) > 200\ {\rm GeV}.
\end{equation}
This will ensure that our signal and background events fall in the same $p_T$ window to do a sensible signal-to-background analysis later in the study. 

\subsection{Cutflow and $b$-tagging Implementation}
The introduction of the full sequence of cuts that we have adopted here requires some justification. In existing $b$-jet analyses that seek to extract chain decays of Higgs bosons from the background, restrictive cuts have been used for ensuring the extraction of a fully hadronic signature. A full description of the cutflow is given in Fig.~\ref{fig:cutflow}.
\vspace*{1em}
\tikzstyle{node} = [rectangle, rounded corners, minimum width=3cm, minimum height=1cm,text centered, draw=black]
\tikzstyle{arrow} = [thick,->,>=stealth]

\begin{figure}[!htb]
\centering

\begin{tikzpicture}[node distance=1.5cm]
\node (cut1) [node, align=center] {Perform fast detector simulation using {\tt Delphes}};
\node (cut2) [node, align=center, below of=cut1] {Perform jet reconstruction and remove jets with $p_T <$ 200 GeV \\ in {\tt fastjet} \cite{Cacciari:2011ma}, with specified clustering algorithm and $\Delta R$};
\node (cut3) [node, align=center, below of=cut2] {Apply double $b$-tagging on the jets};
\node (cut4) [node, align=center, below of=cut3] {When two fat double $b$-tagged jets remain, \\ calculate the invariant mass and compare with $m_H$};

\draw [arrow] (cut1) -- (cut2);
\draw [arrow] (cut2) -- (cut3);
\draw [arrow] (cut3) -- (cut4);

\end{tikzpicture}
\caption{Description for jet clustering, $b$-tagging and selection of jets.}
\label{fig:cutflow}
\end{figure}
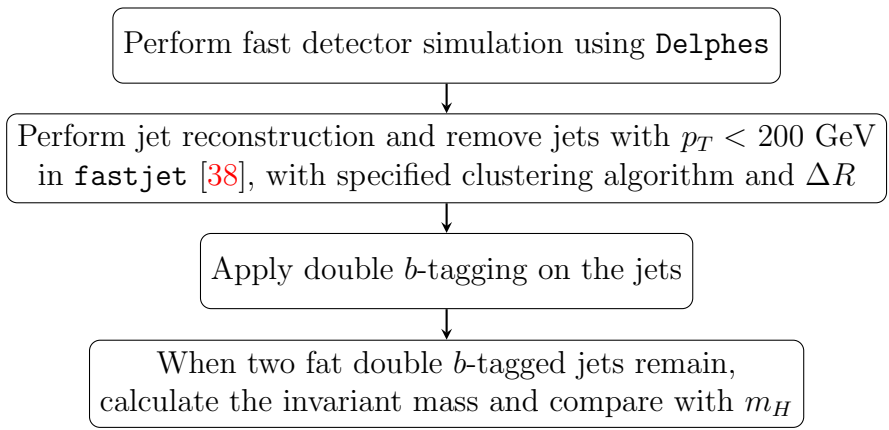

In this paper, we implement a simplified (MC truth informed) double  $b$-tagger. For events clustered using the anti-$k_T$ algorithm (C/A algorithm) with  fixed-$R$ cone size, parton level $b$-(anti)quarks within angular distance $R$ from jets are searched for and if there are two $b$-quarks present within that separation, jets are tagged as double $b$-tagged fat jets as appropriate. When the variable-$R$ approach is used, the size of the tagging cone is taken as the effective size $R_{\rm eff}$ of the jet.

In addition, we account for the finite efficiency of identifying a $b$-jet as well as the non-zero probability that $c$-jets and light-flavour and gluon jets are mistagged as $b$-jets. We apply $p_T$-dependent tagging efficiencies and mistag rates from a {Delphes CMS detector card}\footnote{See {\tt https://github.com/delphes/delphes/blob/master/cards/delphes$_{-}$card$_{-}$CMS.tcl}.}. Note that we have checked that the conclusion remains the same if we use a modified $b$-tagging procedure by replacing the $b$-partons with $b$-hadrons produced after the hadronisation of the $b$-quarks. 

\section{Results}
In this section, we present our results for both the signal and dominant SM backgrounds, first at the parton level and then at the detector level. The dominant backgrounds, such as the QCD $4b$ continuum as well as the $gg,q \bar q \rightarrow t \bar t$ and  $gg,q \bar q \rightarrow Zb\bar b$ channels, are considered for the signal-to-background analysis later in the study.

\subsection{Parton Level Analysis}
Before proceeding with the detector level analysis, we take a look at the parton level information of the events, in order to tweak certain parameters for jet clustering, as well as for sensibly using the selected kinematic cuts. In fact, the $p_T$ of the final state $b$-partons will inform us which value of $\rho$ to use for the variable-$R$ clustering algorithm.
\begin{figure}[htb!]
\centering
\includegraphics[scale=0.4]{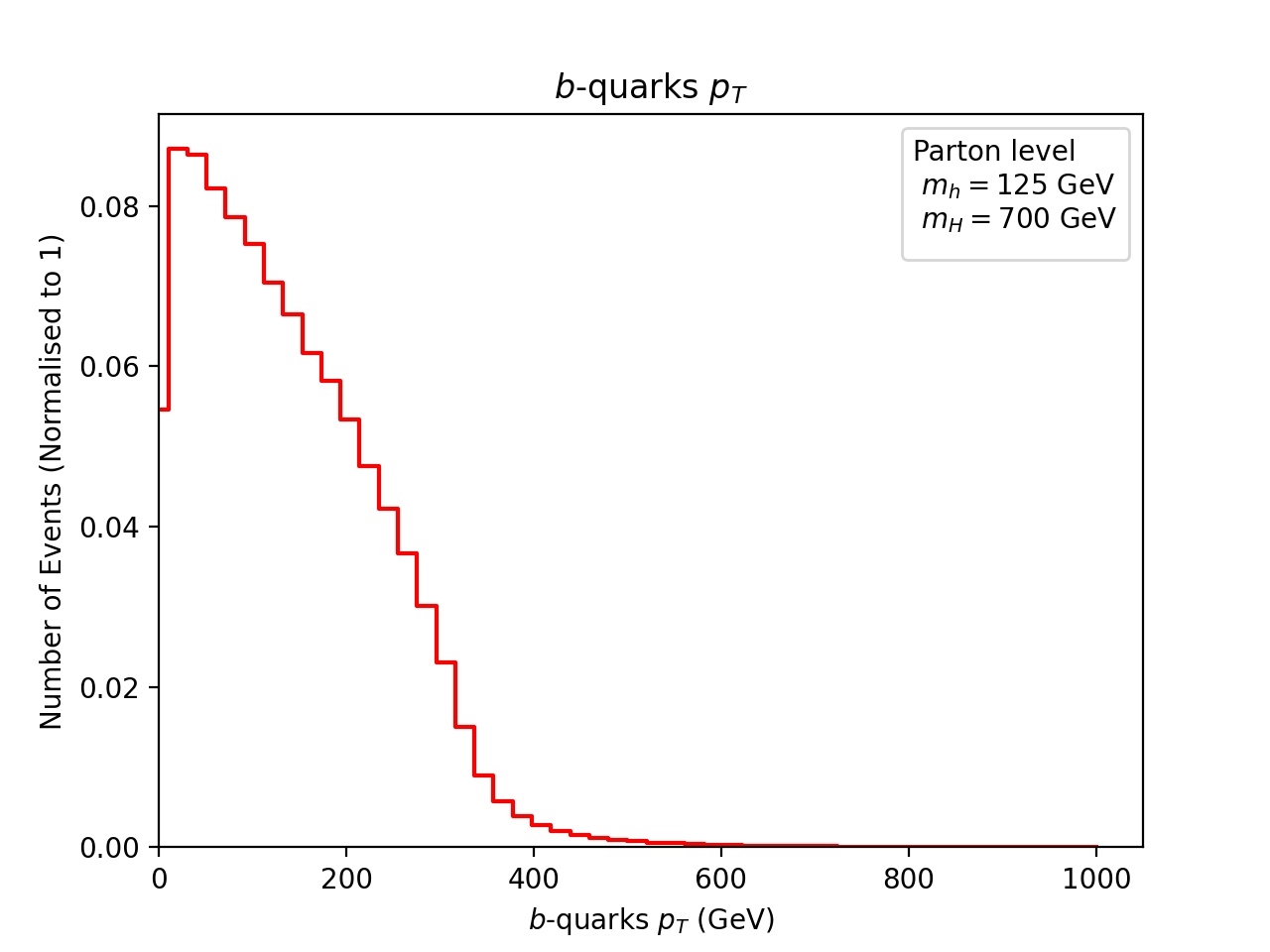}
\includegraphics[scale=0.4]{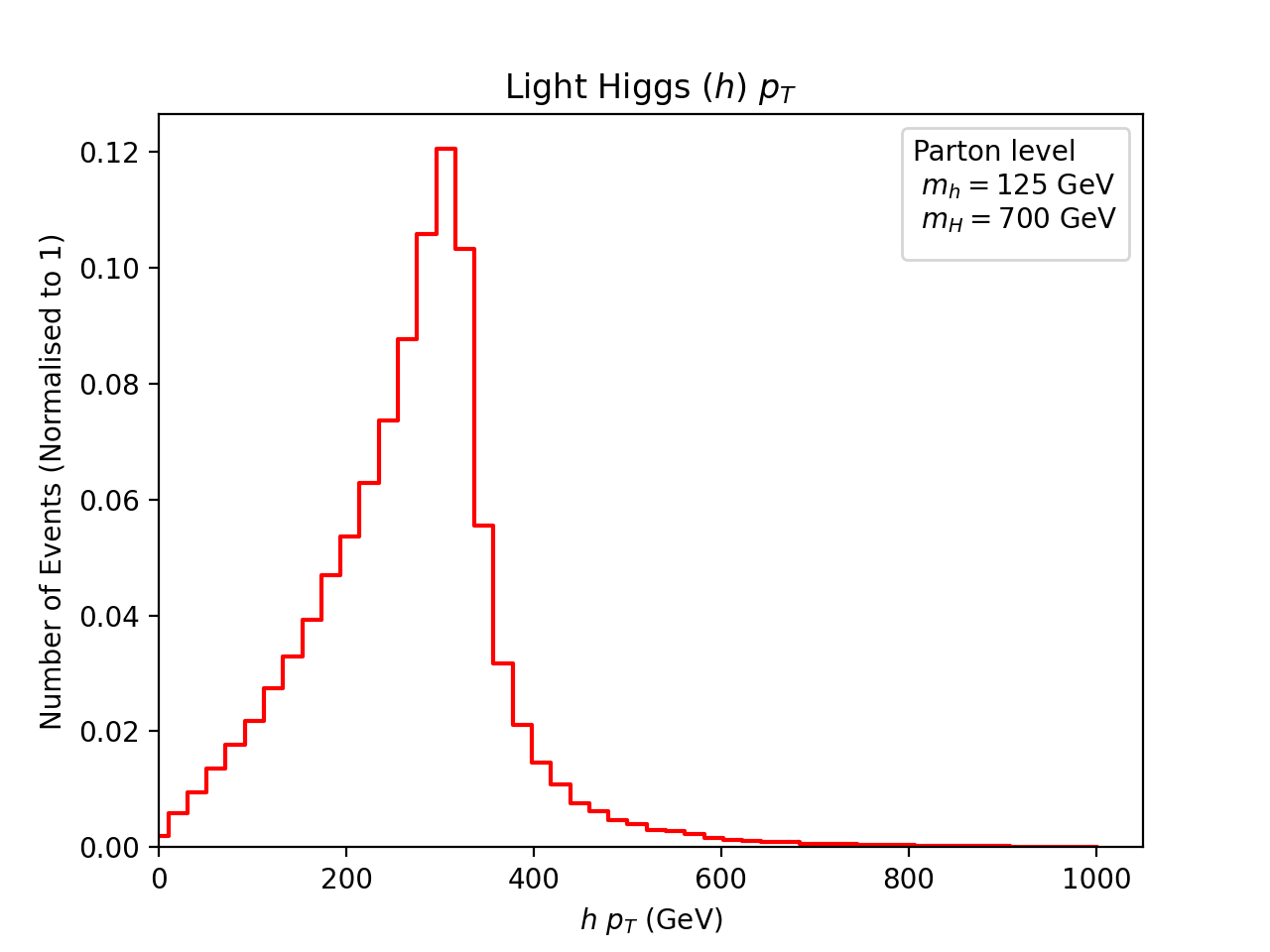}
\caption{Transverse momenta of the final state $b$-quarks (left) and lights Higgses (right) before showering and hadronisation. }
\label{fig:parton}
\end{figure}
\\
From Fig.~\ref{fig:parton} (left), we can see that the final state $b$-quarks have a wide range of momenta, well into $O(10^2)$ GeV. The value of $\rho$, the variable-$R$ specific parameter, is generally chosen to be of the same order of magnitude as the jet $p_T$. However, looking at the $p_T$ distribution of the $b$-quarks, we perform a scan for $\rho$ over the region $[100, 500]$ GeV to find an optimal value. Another point to mention here is that the light Higgs bosons are quite boosted, as seen from Fig.~\ref{fig:parton} (right). The angular separation in the $\eta-\phi$ plane between the pairs of Higgs bosons as well as $b$-quark pairs coming from the same Higgs boson crucially depend on the $p_T$ of the heavier and lighter Higgs bosons. 

\begin{figure}[htb!]
\centering
\includegraphics[scale=0.4]{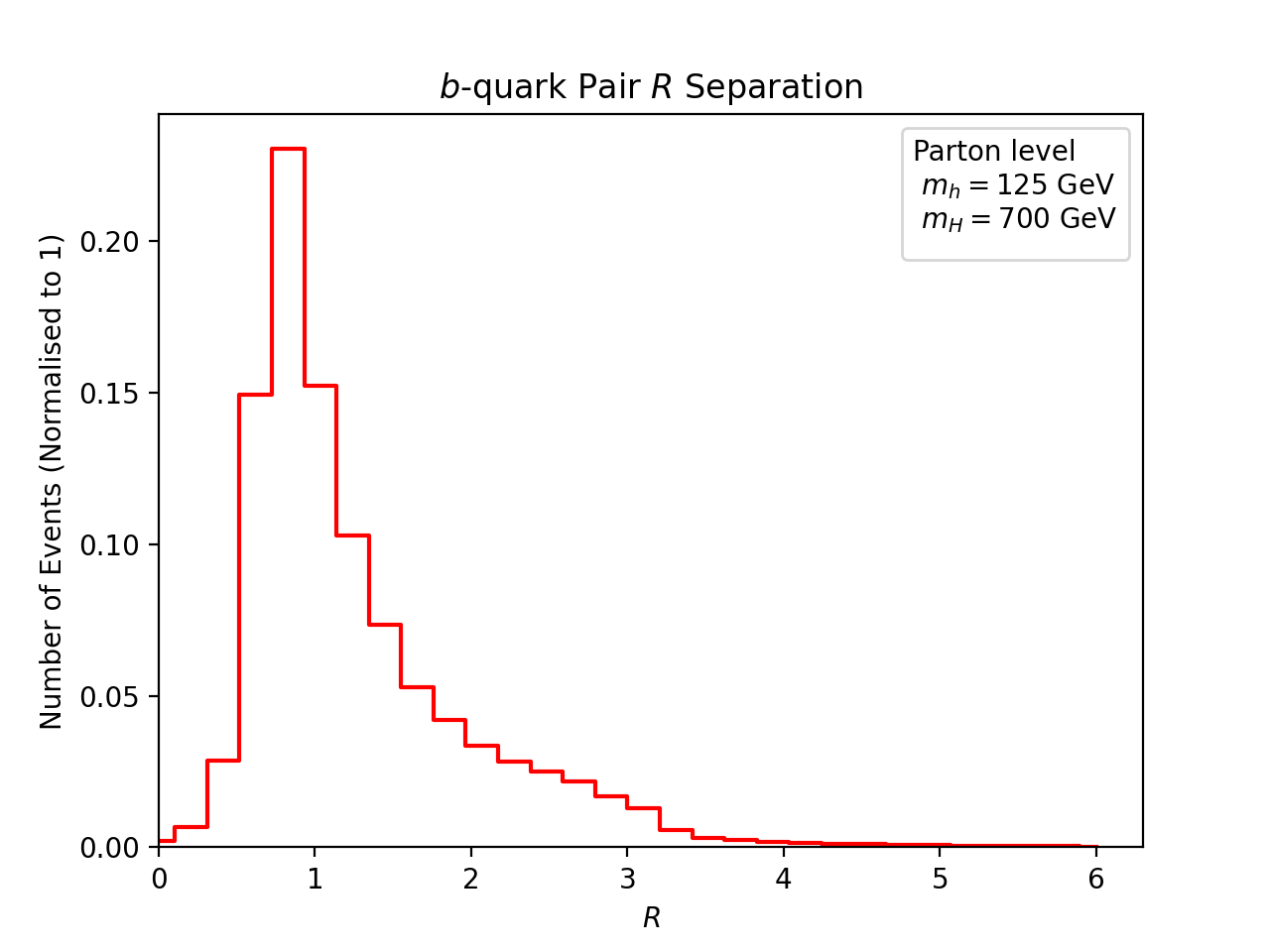}
\includegraphics[scale=0.4]{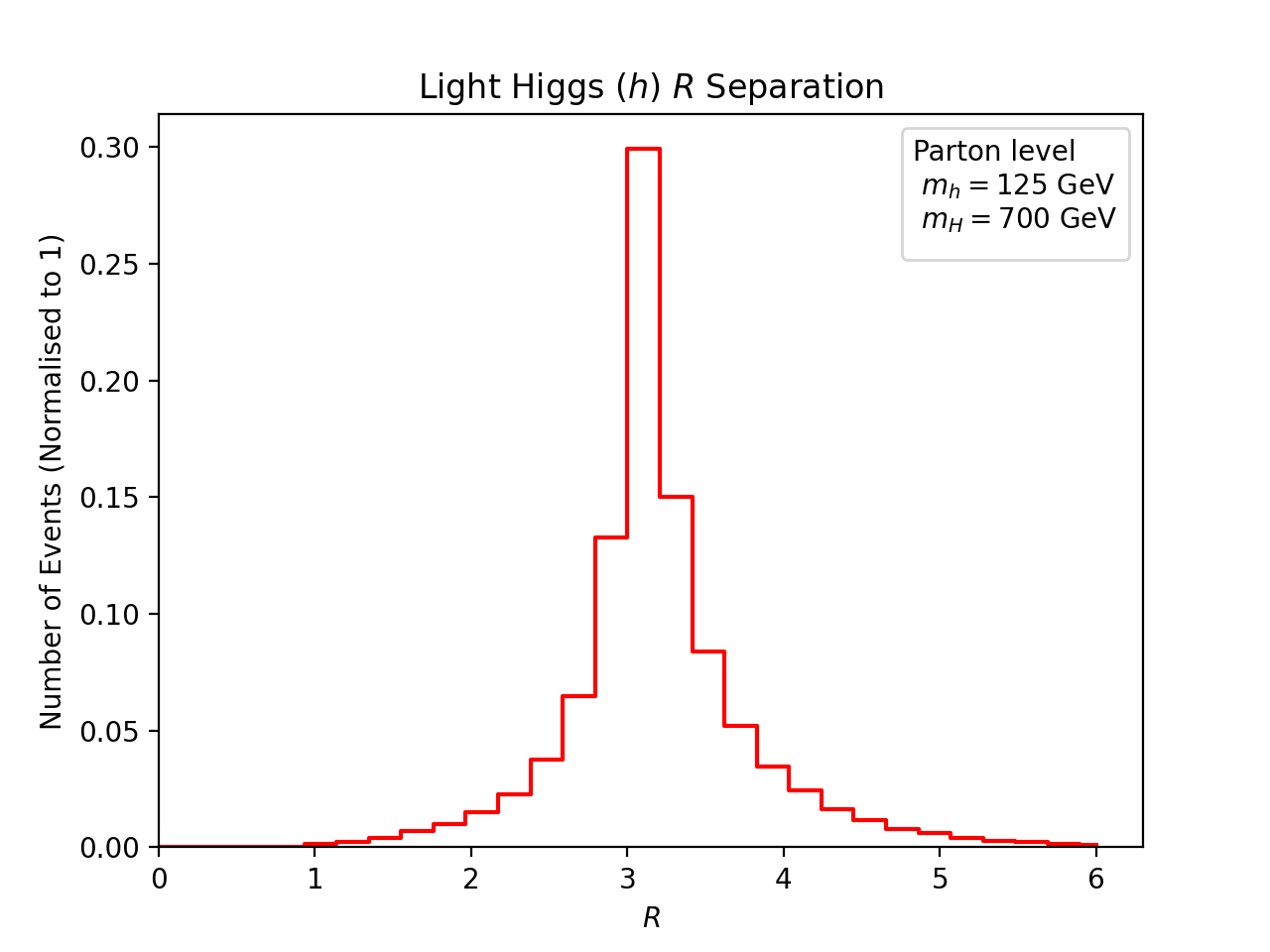}
\caption{$\Delta R$ separation of the  $b \bar b$ pair from a given Higgs (left) and between the two Higgses (right).}
\label{fig:parton_delr}
\end{figure}

In Fig.~\ref{fig:parton_delr}, we see that the two light Higgs bosons are generally always back-to-back, their angular separation peaks being around $\pi$, which implies that the heavier 
Higgs boson is mostly produced at rest. Even though the heavier Higgs boson has very negligible $p_T$, due to the mass configuration of this BP, the two SM-like Higgses have a large momentum transfer from the heavy Higgs boson. The $b$-quarks originating from the lighter Higgs bosons, in contrast, tend to be closer together, i.e., collimated, which is an artefact of boosting. 
Consequently, the resulting jets from these $b$-partons will be close together in detector space. We can exploit this, and instead of trying to lower the values of $R$ in the  jet clustering algorithm to ‘pick out’ and tag all four signal $b$-jets, we can instead use a deliberately large cone in order to capture two fat (and back-to-back) jets, wherein each contains both $b$-quarks coming from the decayed SM-like Higgs boson. 

\subsection{Jet Level Analysis}
\label{sec:jetlevel}
Now, informed by the parton level kinematics of the events, we can proceed to analyse this topology at the jet level. Using the anti-$k_T$ algorithm \cite{Cacciari:2008gp}, we cluster EFlow objects obtained from after the fast detector simulation using Delphes into wide cone jets. We select those jets which have a $p_T>200$ GeV before we proceed to tag them, as described in Fig.~\ref{fig:cutflow}\footnote{We have switched on ISR and MPI  in {\tt Pythia8} to investigate the results for the two types of algorithms in section 4.2.}. 

Here, we compare two different methods of jet clustering for these double $b$-tagged fat jets. Firstly we use a large fixed cone size $R=1.0$ to construct two (nearly) back-to-back fat jets from each $h$ decay, which individually should reveal the mass of the SM-like Higgs boson \footnote{We did optimise the fixed cone size value and $R=1.0$ was found to be the best choice for the reconstruction of mass peaks.}.  Secondly, we do the same but consider the variable-$R$ jet clustering algorithm \cite{Krohn:2009zg}. We optimise the choice of $\rho$ to obtain the best reconstructed resonance mass peaks. For variable-$R$, we use $\rho=300$ with $R_{\rm min}=0.4$ and $R_{\rm max}=2.0$. These values are informed by the $p_T$ scale of the fixed cone $b$-jets and also the aforementioned scan on $\rho$.  

\begin{figure}[htb!]
\centering
\includegraphics[scale=0.6]{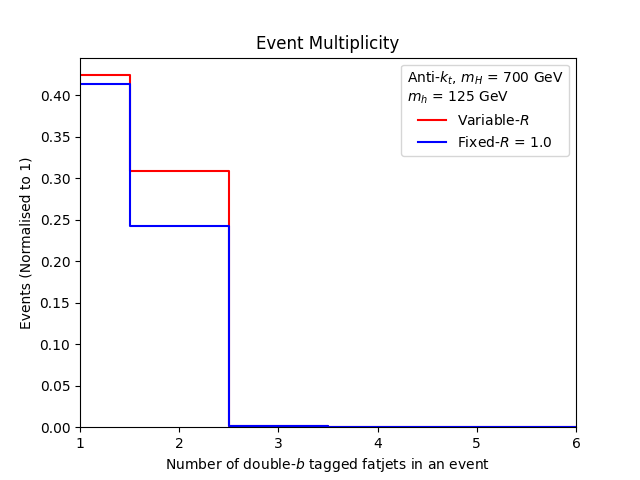}
\caption{The double $b$-tagged fat jets multiplicity distribution for our BP.}
\label{fig:bjet_mult}
\end{figure}

In Fig.~\ref{fig:bjet_mult}, we compare the $b$-jet multiplicity of the signal events for both fixed-$R$ and variable-$R$ algorithms. It is clear from the figure that we obtain more events with double-$b$ tagged fat jets for variable-$R$ than for fixed-$R=1.0$. The presence of more events containing double-$b$ tagged fat jets from the signal allows us to better reconstruct the Higgs resonance peaks in multi-jet mass distributions.

\begin{figure}[htb!]
\centering
\includegraphics[scale=0.5]{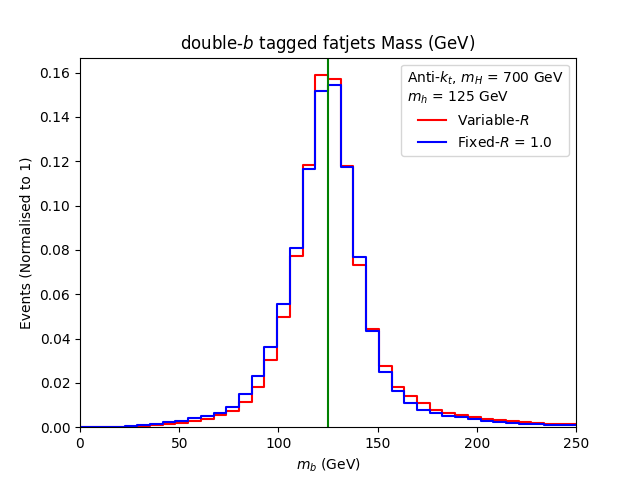}
\includegraphics[scale=0.5]{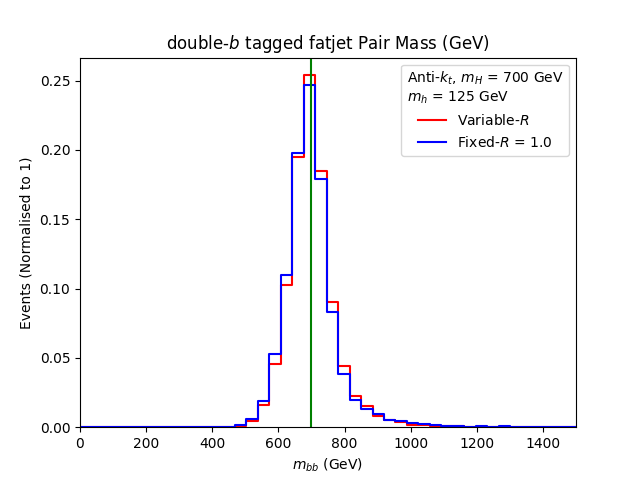}
\caption{Right: The double $b$-tagged fat jets invariant mass $m_h$ for our BP. Left: The two double $b$-tagged fat jets invariant mass $m_H$ for our BP.}
\label{fig:bjet_mass}
\end{figure}

Next, to show the evidence of new physics, we reconstruct the mass of the resonances, namely the light and heavy Higgs bosons. We show the invariant masses of individual double $b$-tagged fat jets  and the pair of double-$b$ tagged fat jets  in Fig.~\ref{fig:bjet_mass}. For $m_h$ mass resonance, we select the average of all double $b$-tagged jets.For $m_H$ resonance, we select events with two double $b$-tagged jets in order to recover heavy Higgs peak. It is evident that the peak of the variable-$R$  algorithm mass distributions is closer to the MC truth value of the corresponding Higgs boson masses, namely $m_h$ = 125 GeV and $m_H$ = 700 GeV.

\begin{figure}[htb!]
\centering
\includegraphics[scale=0.5]{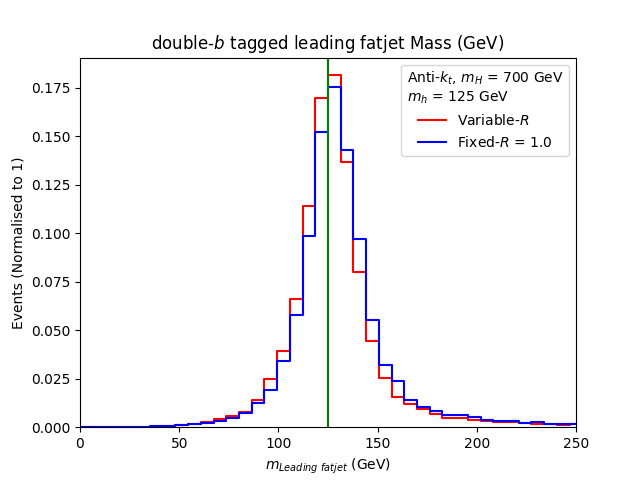}
\includegraphics[scale=0.5]{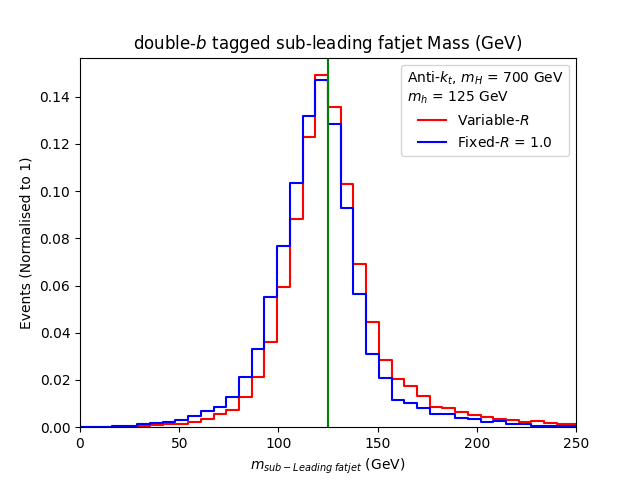}
\caption{Right: The double $b$-tagged leading fat jet invariant mass $m_h$ for our BP. Left: The double $b$-tagged sub-leading fat jet invariant mass $m_h$ for our BP.}
\label{fig:bjet_mass_lead_sub}
\end{figure}
For completeness, we also present mass distributions for the  leading and subleading fat jets  (double $b$-tagged) in Fig.~\ref{fig:bjet_mass_lead_sub}. The same behaviour can be seen here with variable-$R$ jet algorithm results being more aligned towards the corresponding MC truth value of the light Higgs boson mass. As a next step,  we look at signal-to-background rates to compare the two jet reconstruction algorithms mentioned in the paper in this respect. 

\subsection{Signal-to-background Analysis}
Here, we describe the performance of our final cuts used in extracting the signal from the backgrounds and compute the final significances in presence of both MPI and PU effects.
\subsubsection{Signal-to-background Analysis with MPIs}
In order to quantify the performance of the variable-$R$ algorithm against the fixed-$R$ one, we calculate signal-to-background rates and signal significances for the aforementioned two choices of integrated luminosity. To carry out this exercise, we apply the additional selection procedure 

\begin{figure}[!ht]
\centering

\begin{tikzpicture}[node distance=1.5cm]
\node (cut1) [node,align=center] {Select events that contain exactly two double \\ $b$-tagged fat jets};
\node (cut2) [node, align=center,below of=cut1] {Select event if the double $b$-tagged fat jets \\ invariant mass $m_h$ falls under [100,150] GeV range};
\node (cut3) [node,align=center, below of=cut2] {Select event if the two double $b$-tagged fat jets \\ invariant mass $m_H$ falls under [650,750] GeV range};

\draw [arrow] (cut1) -- (cut2);
\draw [arrow] (cut2) -- (cut3);
\end{tikzpicture}
\caption{Additional event selection used to compute the final signal-to-background rates.}
\label{fig:selection}
\end{figure}
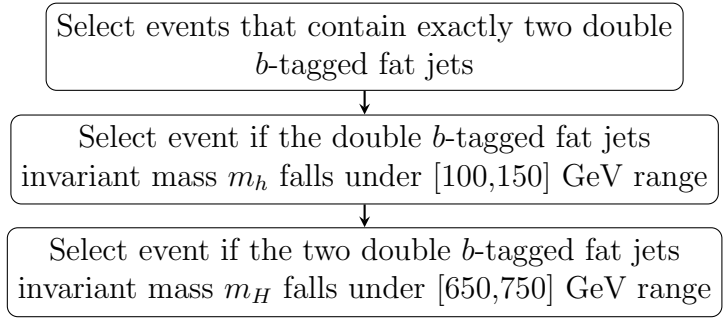

described in Fig.~\ref{fig:selection}. 

The event rates ($N$) for the various processes is given by: 
\begin{equation}
N \ = {\rm Cross~section}~(\sigma) \times {\rm Luminosity}~(\mathcal{L}).
\end{equation}

From Tabs.~\ref{tab:signalbackground2} and \ref{tab:signalbackground3}, it is evident that $pp\rightarrow b \bar b b \bar b$ is the dominant background process followed by $pp\rightarrow t \bar t$ and $pp\rightarrow Z b \bar b$. Upon using the two values of integrated luminosities ${\cal L}=$ $140$ fb$^{-1}$ and $300$ fb$^{-1}$, the next step is to calculate the significance rates ($\Sigma$) as a function of signal ($S$) and background ($B$) rates, which is given by:
\begin{equation}
\Sigma = \frac{N(S)}{\sqrt{N(B_{b\bar{b}b\bar{b}})+N(B_{t\bar{t}})+N(B_{Zb\bar{b}})}}.
\end{equation}
Tab.~\ref{tab:signalbackground4} contains the significances for both choices of the jet clustering algorithm without and with QCD $K$-factors. The QCD $K$-factors describe the ratio between the leading and higher order cross sections. We have used $K$ = 2 (at NNLO level) for the signal \cite{Ravindran:2003um,Harlander:2002wh}, $K$ = 1.5 (at NLO level)  for $pp \rightarrow b\bar{b}b\bar{b}$ \cite{Greiner:2011mp}, $K$ = 1.4 (at NLO level)  for $pp \rightarrow t\bar{t}$ \cite{SM:2010nsa} and $K$ = 1.4 (at NLO level)   for $pp \rightarrow Z b\bar{b}$ \cite{FebresCordero:2009xzo}). It is clear that the variable-$R$ approach is more efficient compared to the fixed-$R$ method. The conclusion remains the same even after we take into account a typical 10\% effect of systematic uncertainties in our calculation of signal significances.

We also present the significances for both choices of jet clustering algorithms without and with QCD $K$-factors using Trimming grooming techniques \cite{Krohn:2009th} to mitigate the effect of ISR and MPI in Tab.~\ref{tab:signalbackground5}. We have used default CMS values for $R_{Trim} = 0.2$ and $p_{T_{FracTrim}} = 0.05$ taken from the Delphes CMS detector card. It is again clear that the variable-$R$ approach is more efficient compared to the fixed-$R$ method and our conclusions still hold even after the jets are groomed (one can always use other grooming techniques such as filtering \cite{Butterworth:2008iy}, pruning \cite{Ellis:2009su}, mass-drop \cite{Butterworth:2008iy}, modified mass-drop \cite{Dasgupta:2013ihk} and soft drop \cite{Larkoski:2014wba}, however, this is beyond the scope of this paper).
\begin{table}[!h]
\scalebox{1.0}{
\begin{tabular}{|l|l|l|l|l|}
    \hline
    \multirow{2}{*}{Process} &
      \multicolumn{1}{c}{Variable-$R$}\vline &
      \multicolumn{1}{c}{$R=1.0$ } \vline \\ \cline{2-3}
    & $m_h$ = 125 GeV, $m_H$ = 700 GeV&$m_h$ = 125 GeV, $m_H$ = 700 GeV  \\
    \hline
    $pp\to H\to hh  \rightarrow b\bar{b}b\bar{b}$ &  147.56&  104.874  \\
    \hline
    $pp \rightarrow t\bar{t}$ & 166.633 & 111.088   \\
    \hline
    $pp \rightarrow b\bar{b}b\bar{b}$ &592.336&435.139   \\
    \hline
    $pp \rightarrow Z b\bar{b}$ &0.067&0.063\\
    \hline
  \end{tabular}
}
\caption{\label{tab:signalbackground2} Event rates of signal and backgrounds for ${\cal L}=$ $140$ fb$^{-1}$  upon enforcing the initial cuts plus the mass selection criteria of Fig.~\ref{fig:selection} for the two jet reconstruction procedures.}
\end{table}

\begin{table}[!h]
\scalebox{1.0}{
\begin{tabular}{|l|l|l|l|l|}
    \hline
    \multirow{2}{*}{Process} &
      \multicolumn{1}{c}{Variable-$R$}\vline &
      \multicolumn{1}{c}{$R=1.0$ } \vline \\ \cline{2-3}
    & $m_h$ = 125 GeV, $m_H$ = 700 GeV&$m_h$ = 125 GeV, $m_H$ = 700 GeV  \\
    \hline
    $pp\to H\to hh  \rightarrow b\bar{b}b\bar{b}$ &  316.2&  224.73  \\
    \hline
    $pp \rightarrow t\bar{t}$ & 357.071 & 238.047  \\
    \hline
    $pp \rightarrow b\bar{b}b\bar{b}$ &1269.292&932.441  \\
    \hline
    $pp \rightarrow Z b\bar{b}$ &0.145&0.135\\
    \hline
  \end{tabular}
}
\caption{\label{tab:signalbackground3} Event rates of signal and backgrounds for ${\cal L}=$ $300$ fb$^{-1}$  upon enforcing the initial cuts plus the mass selection criteria of Fig.~\ref{fig:selection} for the two jet reconstruction procedures.}
\end{table}

\begin{table}[!htb]
\begin{center}
\scalebox{1.0}{
\begin{tabular}{ |c|c|c| }
 \hline
 & Variable-$R$  & $R=1.0$     \\
 \hline
${\cal L}=$ $140$ fb$^{-1}$ &5.355  & 4.487   \\
 \hline
${\cal L}=$ $300$ fb$^{-1}$ & 7.840& 6.568   \\
\hline
\end{tabular}
}
\scalebox{1.0}{
\begin{tabular}{ |c|c|c| }
 \hline
 & Variable-$R$  & $R=1.0$     \\
 \hline
${\cal L}=$ $140$ fb$^{-1}$ &8.810  & 7.377   \\
 \hline
${\cal L}=$ $300$ fb$^{-1}$ & 12.897& 10.799   \\
\hline
\end{tabular}
}
\caption{\label{tab:signalbackground4} Left panel: Final $\Sigma$ values calculated upon enforcing the initial cuts plus the mass selection criteria 
 of Fig.~\ref{fig:selection}
for the two jet reconstruction procedures. Right panel: The same in presence of $K$-factors.}
\end{center}
\end{table}

\begin{table}[!htb]
\begin{center}
\scalebox{1.0}{
\begin{tabular}{ |c|c|c| }
 \hline
 & Variable-$R$  & $R=1.0$     \\
 \hline
${\cal L}=$ $140$ fb$^{-1}$ &5.753  & 4.861   \\
 \hline
${\cal L}=$ $300$ fb$^{-1}$ & 8.421& 7.116   \\
\hline
\end{tabular}
}
\scalebox{1.0}{
\begin{tabular}{ |c|c|c| }
 \hline
 & Variable-$R$  & $R=1.0$     \\
 \hline
${\cal L}=$ $140$ fb$^{-1}$ &9.513  & 8.022   \\
 \hline
${\cal L}=$ $300$ fb$^{-1}$ & 13.926& 11.743   \\
\hline
\end{tabular}
}
\caption{\label{tab:signalbackground5} Left panel: Final $\Sigma$ values calculated upon enforcing the initial cuts plus the mass selection criteria 
 of Fig.~\ref{fig:selection}
for the two jet reconstruction procedures using Trimming grooming techniques. Right panel: The same in presence of $K$-factors.}
\end{center}
\end{table}

\subsubsection{Signal-to-background Analysis with Pile-Up}
As a final exercise, we want to check the performance of the two clustering algorithms used in this paper to reconstruct jets with Pile-Up (PU).  As mentioned previously, to perform such a study one needs to apply proper detector simulation using  {\tt Delphes}. Specifically, generated events after hadronisation are passed through a {\tt Delphes} CMS PU card\footnote{See
 {\tt https://github.com/recotoolsbenchmarks/DelphesNtuplizer/blob/master/cards/CMS$_{-}$PhaseII$_{-}$200PU
$_{-}$Snowmass2021$_{-}$v0.tcl{\#}L1039-L1067.}}. To generate the PU simulations, we have used {\tt Pythia8}. Mixing of these PU events with the signal events is then done with $<N_{\rm PU}>$  = 50 for each hard scattering. Next, {\tt FastJet}  is implemented for both the variable-$R$ and anti-$k_T$ (with $R=1.0$)  algorithms within the same card, to finally output jet information into a {\tt Root file}. We finally carry out the analysis through a {\tt Root} macro code using the same cutflow described in Section 3.2 in presence of the additional selection procedure described in Fig.~\ref{fig:selection}.

We again calculate the signal-to-background rates, and consequent significances, in presence of the usual luminosities, specifically, for the purpose of comparing the performance of the variable-$R$ jet clustering algorithm against the anti-$k_T$ one with fixed $R=1.0$ in extracting the signal from the dominant backgrounds. The event rates ($N$) (described by Eq.~(11)) for the various processes are given in Tabs.~\ref{tab:signalbackground6} and \ref{tab:signalbackground7}.
\begin{table}[!h]
\scalebox{1.0}{
\begin{tabular}{|l|l|l|l|l|}
    \hline
    \multirow{2}{*}{Process} &
      \multicolumn{1}{c}{Variable-$R$}\vline &
      \multicolumn{1}{c}{$R=1.0$ } \vline \\ \cline{2-3}
    & $m_h$ = 125 GeV, $m_H$ = 700 GeV&$m_h$ = 125 GeV, $m_H$ = 700 GeV  \\
    \hline
    $pp\to H\to hh  \rightarrow b\bar{b}b\bar{b}$ &  76.655&  55.239  \\
    \hline
    $pp \rightarrow t\bar{t}$ & 111.088 & 166.633   \\
    \hline
    $pp \rightarrow b\bar{b}b\bar{b}$ &423.748&282.498   \\
    \hline
    $pp \rightarrow Z b\bar{b}$ &0.0180&0.0270\\
    \hline
  \end{tabular}
}
\caption{\label{tab:signalbackground6} Event rates of signal and backgrounds for ${\cal L}=$ $140$ fb$^{-1}$  upon enforcing the initial cuts plus the mass selection criteria of Fig.~\ref{fig:selection} for the two jet reconstruction procedures.}
\end{table}

\begin{table}[!h]
\scalebox{1.0}{
\begin{tabular}{|l|l|l|l|l|}
    \hline
    \multirow{2}{*}{Process} &
      \multicolumn{1}{c}{Variable-$R$}\vline &
      \multicolumn{1}{c}{$R=1.0$ } \vline \\ \cline{2-3}
    & $m_h$ = 125 GeV, $m_H$ = 700 GeV&$m_h$ = 125 GeV, $m_H$ = 700 GeV  \\
    \hline
    $pp\to H\to hh  \rightarrow b\bar{b}b\bar{b}$ &  164.260&  118.371  \\
    \hline
    $pp \rightarrow t\bar{t}$ &238.047  & 357.071  \\
    \hline
    $pp \rightarrow b\bar{b}b\bar{b}$ &908.032&605.354  \\
    \hline
    $pp \rightarrow Z b\bar{b}$ &0.038&0.0580\\
    \hline
  \end{tabular}
}
\caption{\label{tab:signalbackground7} Event rates of signal and backgrounds for ${\cal L}=$ $300$ fb$^{-1}$  upon enforcing the initial cuts plus the mass selection criteria of Fig.~\ref{fig:selection} for the two jet reconstruction procedures.}
\end{table}

\begin{table}[!htb]
\begin{center}
\scalebox{1.0}{
\begin{tabular}{ |c|c|c| }
 \hline
 & Variable-$R$  & $R=1.0$     \\
 \hline
${\cal L}=$ $140$ fb$^{-1}$ &3.314  & 2.606   \\
 \hline
${\cal L}=$ $300$ fb$^{-1}$ & 4.851& 3.815   \\
\hline
\end{tabular}
}
\hspace{1mm}
\scalebox{1.0}{
\begin{tabular}{ |c|c|c| }
 \hline
 & Variable-$R$  & $R=1.0$     \\
 \hline
${\cal L}=$ $140$ fb$^{-1}$ &5.450  & 4.309   \\
 \hline
${\cal L}=$ $300$ fb$^{-1}$ & 7.978& 6.309   \\
\hline
\end{tabular}
}
\caption{\label{tab:signalbackground8} Left panel: Final $\Sigma$ values calculated calculated upon enforcing the initial cuts plus the mass selection criteria of Fig.~\ref{fig:selection} for the two jet reconstruction procedures. Right panel: The same in presence of  $K$-factors.}
\end{center}
\end{table}

Finally, Tab.~\ref{tab:signalbackground8} contains the final significance rates (as per Eq.~(12)), again  without and with $K$-factors. It is clear that the variable-$R$ approach is again more efficient compared to the fixed-$R$ method even with PU effects  added.  

\section{Summary and Conclusions}
In this paper, we have studied the performance of two different kinds of  jet clustering algorithms at the LHC in accessing BSM signals induced by the cascade decays of a heavy Higgs boson $H$ (with a mass of 700 GeV) into a pair of SM-like Higgs states, $hh$. Given the mass difference between the two Higgs masses involved, the lighter Higgs bosons are produced with a large boost, so that their decays products, namely, a pair of $b$-quarks in our study, become highly collimated. We, therefore, reconstruct these events into two fat jets and perform a double $b$-tagging on these. For illustrative purposes, a 2HDM-II setup was assumed, by adopting a BP over its parameter space fully compliant with both theoretical and experimental constraints.

The two different kinds of jet clustering algorithms are a variable-$R$ one (where the cone size is not fixed but rather adapts to  the resonant kinematics of the signal) and a more standard one, with a fixed cone size ($R=1$).   These are used to reconstruct the mass of the lighter (SM-like) Higgs boson twice. Further, we select those events where a pair of such double $b$-tagged fat jets exist, in which total invariant mass reproduces the heavy Higgs boson mass. Through a cut-based signal-to-background analysis, we further  find that the variable-$R$ method not only provides better reconstructed peaks of both Higgs boson masses, compared to the traditional algorithm, but also improves the signal-to-background ratio, which in turn results in higher signal significances at the LHC (altogether leading to potential discovery at both Run 2 and 3 of the LHC).  Thus, we advocate the use of the former in establishing $pp\to H\to hh\to b\bar b b\bar b$ events in boosted topologies, in line with similar  results previously obtained for the case of the same channel and different mass spectra yielding four slim $b$-jets.
Finally, note that we have used the anti-$k_T$ algorithm as representative of the fixed cone size kind throughout but results are the same for the C/A jet clustering algorithm.

\section*{Acknowledgements}
\noindent
SM is supported in part through the NExT Institute and the STFC Consolidated Grant ST/L000296 /1. SJ is partially funded by DISCnet studentship. The work of AC is funded by the Department of Science and Technology, Government of India, under Grant No. IFA18-PH 224 (INSPIRE Faculty Award). We all thank Claire H. Shepherd-Themistocleous and Emmanuel Olaiya for useful discussions.  SJ
acknowledge the use of the IRIDIS5 High Performance Computing Facility, and associated support services at the University of Southampton, in the completion of this work.

\end{document}